\documentclass[aps,prb,twocolumn,showpacs,floatfix]{revtex4}
\usepackage{graphicx}

\begin{document}
\title{ Ab-initio Phonon Calculations for the layered compound TiOCl}

\author{Leonardo Pisani and
Roser Valent{\'\i}}

\affiliation{ 
Institut f{\"u}r Theoretische Physik, Universit{\"a}t Frankfurt, 
D-60054 Frankfurt, Germany}

\date{\today}

\begin{abstract}
\footnotesize{
We present first-principles frozen-phonon calculations for the
 three Raman-active $A_g$ modes in the spin-1/2
layered TiOCl system within two different well-known approaches:
the local density approximation (LDA) and the so-called LDA+U
approximation.  We observe that the inclusion of electron correlation
in a mean-field level as implemented in the LDA+U leads to a better overall
 agreement with experimental results. We also discuss the implications of the two approaches 
 on the physics of TiOCl.} 

\end{abstract}
\maketitle

{\it Introduction}.-
 The layered quantum spin system TiOCl has been recently a subject of 
 debate due to its anomalous properties at moderate to low temperatures.
Susceptibility measurements\cite{Seidel_03}
show a kink at T$_{c2}$=94 K and an exponential drop at T$_{c1}$=66 K
indicating the opening of a spin gap which has been interpreted
as a spin-Peierls phase transition\cite{Seidel_03}. LDA+$U$ calculations
performed by these authors predicted the system to behave as a spin-chain along the $b$ axis.
X-ray diffraction measurements below T$_{c1}$=66 K report\cite{vanSmaalen04}
 the presence of superlattice reflections at $(h,k+1/2,l)$
which denotes a doubling of the unit cell along the $b$ axis
and has been interpreted as a confirmation that the system
undergoes a spin-Peierls phase transition below T$_{c1}$.
Still unresolved is the nature of the kink at T$_{c2}$.
 The observation in Raman and infrared spectroscopy
of  phonon anomalies \cite{lemmens_03_2} above T$_{c2}$ 
as well as
temperature-dependent g-factors and linewidths in ESR \cite{Kataev_03}
led to the proposal that the system
is subject  
 to  competing lattice, spin, and orbital
degrees of freedom.  Recent heat capacity measurements\cite{Hemberger_2005}
 seem to strongly support the idea of existence of possible lattice and/or
 orbital fluctuations
 above T$_{c2}$. \par
  In view of the above discussion, it is of great importance
 to understand the behavior of the phonons in
 this system as well as the possibility of  coupling 
 to orbital and spin degrees of freedom.
  In a previous communication\cite{saha_dasgupta_04}  
 we presented  
 first-principles results on the electronic
 properties of   a few slightly distorted
 structures for TiOCl
 according to the various A$_g$-Raman active  phonon modes expected for the $Pmmn$
space group assigned to TiOCl. The idea in ref. \cite{saha_dasgupta_04} 
was to investigate the orbital occupancy
  as a function of lattice distortion, which could simulate the phonon-orbital interaction.
 The phonon  modes 
 and the corresponding
eigenvectors  in ref. \cite{saha_dasgupta_04} were obtained by considering  a shell model\cite{kovaleva}.

 In the present work, we abandon the parametrized  procedure implicit
 in the shell model   and we calculate the
 Raman-active $A_g$ phonon modes in TiOCl within a first principles frozen phonon approach.
 From group theoretical analysis, the allowed $A_g$  modes in the $Pmmn$
symmetry define ion displacements along the $c$-axis of the crystal. 
 One important  result of our calculations is that the consideration of correlations
 in this system is of crucial importance in order to get an accurate description
 of the phonons. 
\vspace{0.cm}

{\it First Principles Frozen Phonon approach}.-
 In order to calculate  phonon frequencies within an {\it ab initio}
scheme, two different methods are generally  employed:
 the {\it energy surface}  method \cite{Cohen_89} and the {\it atomic
force}  method \cite{Kouba_99}.  Within  both methods  the harmonic
potentials underlying the lattice vibrations are constructed by selecting frozen-in
distorted  structures   and then performing a  Density Functional
 Theory (DFT) band-structure
calculation. While in the first approach,   energy values are needed to fit the energy
surface  to a bilinear form,  in the second approach 
the atomic forces acting on the displaced ions are computed to fit the harmonic forces. 
 
In both methods the corresponding dynamical matrix  is calculated
according to the following procedure.
Within the Born-Oppenheimer approximation,  the electronic energy hypersurface 
 is represented
as a  function of the titanium, oxygen and  chlorine coordinates along the c-axis.  For
small atomic displacements we  approximate the surface by a Taylor expansion up to
 second order 
around its minimum (harmonic approximation). Since the   minimum satisfies
the equilibrium condition, the surface is described by the following  bilinear form:
\begin{equation}
E-E_0= {1 \over 2}\; \sum_{ij} z_{i} \; K_{ij} \;  z_{j}, \;\;\;
i,j=1,2,3
 \label{bilform}
  \end{equation}
%%%%%%%%%%%%%%%%%%%%%%%%%%%%%%%%%%
 where $z_{i}$ are the c-axis coordinates of the ions $(1\equiv \mbox{titanium},2\equiv
\mbox{oxygen}, 3\equiv \mbox{chlorine})$.  The equation of motion for the $i$-th atom with mass $M_i$ is
then \begin{equation} M_i {d^2 z_i \over dt^2}=-  \frac{\partial E}{\partial z_i}=-\sum_{j}
K_{ij} z_{j}.\label{equmot} \end{equation}
 Seeking for  running-wave solutions of the form:
$ z_i(t)={\xi_i \over \sqrt{M_i} } e^{-i \omega t} \label{zzeta}$
 we are left with the eigenvalue problem $ \omega^2 \xi_i =
\sum_{j} {K_{ij} \over \sqrt{M_i M_j} }\xi_j $ where the matrix
$D_{ij}={K_{ij} \over \sqrt{M_i M_j} }$ is the well-known dynamical matrix.

Within the energy surface method,  Eq.\ \ref{bilform} is exploited to obtain the  matrix $D_{ij}$, while 
within the atomic force method, the left-hand side of Eq.\ \ref{equmot} is used.

 In this work we will make use of both methods within two different
 approximated forms for the exchange-correlation potential in the DFT scheme,
namely,
  the Local Density Approximation (LDA)  \cite{Perdew_92} 
 and the so-called LDA$+U$ approximation \cite{Anisimov_97}.
 For the LDA+$U$ calculations, we considered the implementation, $E^U$ 
 proposed by  M.T.Czyzyk and G.A.Sawatzky\cite{Czyzyk_94}
 (named in the literature "AMF", Around Mean Field)
which differs with respect to the original Anisimov {\em et al.}  energy functional \cite{Anisimov_91}
in that it takes into account  spin degrees of freedom explicitly  for all  electrons in the system.
Notably, the average occupancy of one $d$ orbital $n^0=\frac{1}{2(2l+1)}
\sum_{m\sigma} n_{m\sigma}$ is now replaced by the spin dependent counterpart $n^0_{\sigma}=\frac{1}{2l+1}
\sum_{m} n_{m\sigma}$ and the starting energy functional is  now the spin polarized LSDA functional, $E^S$
\begin{eqnarray}
E^U = E^{S}+{1 \over 2}\sum_{m,m',\sigma} U (n_{m\sigma}-n^0_{\sigma}) 
                                         (n_{m'-\sigma}-n^0_{-\sigma}) \nonumber \\
 +{1 \over 2}\sum_{m\neq m',\sigma}(U-J)(n_{m\sigma}-n^0_{\sigma})(n_{m'\sigma}-n^0_{\sigma}) .
\end{eqnarray}

Calculations have been  performed using the full-potential linearized augmented
plane-wave code WIEN2k\cite{WIEN2k}.
 While the {\it atomic force} method is  more efficient  to obtain
 the dynamical matrix, a successful implementation of this method within
 LDA+$U$ is not yet reliable.  Here we will present results on the 
  structural properties of TiOCl within LDA  considering the {\it atomic force}
 method and within  LDA$+U$ by means of  the {\it total energy} method.
\vspace{0.cm}

{\it Density Functional Calculations}.-
In our calculations, 
  the expansion of the wave functions included 1207 LAPW's\cite{Andersen_75} ($RK_{max}=7$, including 12 Local Orbitals)   and  
the muffin tin radii were chosen to be 1.7 a.u. for
Titanium, 1.5 a.u. for  Oxygen and 2.00 a.u. for Chlorine.  Expansion in spherical
harmonics for the radial wave functions were taken up to $l=10$.  Charge densities and
potentials were represented by spherical harmonics up to $L=6$ whereas in the interstitial
region  they  were expanded in a Fourier series with 4500 stars.  For Brillouin-zone (BZ)
integrations  a  500 ${\bf k}$-point mesh was used  yielding  60  ${\bf
k}$ points in the irreducible wedge and use of the modified tetrahedron method was made \cite{Blochl_94}.  

Concerning the  LDA+$U$ parameters $U$ and $J$ we have used the 
values of 4 eV and 1 eV respectively. Here we have considered a larger
 value of $U$ than in the calculation in ref.\cite{saha_dasgupta_04}(U=3.3eV)
  in line with a Hartree-Fock calculation of on-site Coulomb interaction
for transition metal oxides \cite{mizo_96}.

For a consistent phonon calculation we need to have a well defined 
 {\it ab initio} equilibrium structure which corresponds to the minimum of
 the Taylor expansion Eq.\ \ref{bilform}.
Therefore we   analyze the structure of TiOCl  
by performing an atomic position optimization 
and (within LDA) a unit cell volume optimization.   In the volume optimization,  
 the  $a:b:c$ ratios were kept constant and equal to the experimental values.
For each total energy  calculation we allow  the ions to relax  to
their zero-force positions along the $c$-axis by simulating the dynamics
of  the damped
Newton method.  For the time evolution of the three atomic coordinates,  the method is
based on the finite difference equation  
$z^{\tau+1}_i=z^{\tau}_i+\eta (z^{\tau}_i-z^{\tau-1}_i)+\delta F^{\tau}_i$, 
 where $z^{\tau}_i$ and $F^{\tau}$ are the coordinate and the force on
the $i$-th atom at time step $\tau$. Damping and speed of motion are controlled by
the two parameters $\eta$ and $\delta$ respectively.

%%%%%%%%%%%%%%%%%%%%%%%%%%%%%%%%%%%%%%%%%%%%%%%%%%%%%%%%%%%%%%%%%%%%%%%%%

\begin{figure}
\includegraphics[width=7cm,keepaspectratio]{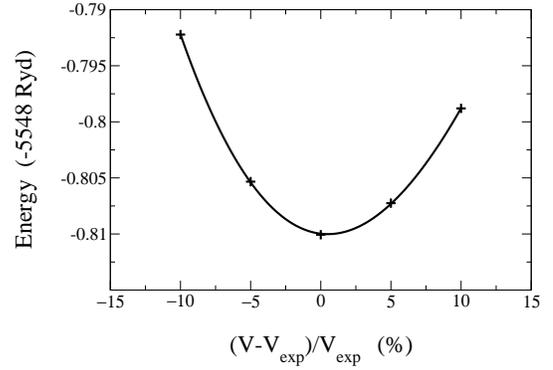}
\caption{Volume optimization for TiOCl within LDA with the ratio $a:b:c$ fixed
to the experimental one.  }
\label{volopt}
\end{figure}

%%%%%%%%%%%%%%%%%%%%%%%%%%%%%%%%%%%%%%%%%%%%%%%%%%%%%%
 
 In Fig.\ \ref{volopt} the total  energy  is displayed versus the volume
variation with respect to the experimental reference value.  The
 continuous line represents a fit to the  Murnaghan equation of state \cite{murn_44}
 from which we  extract a bulk modulus value of 59.5 GPa.
The curve shows that the optimal volume is in  good agreement  with the experimental one,  
in contrast with results obtained for high-T$_c$ compounds\cite{Kouba_99}.

%%%%%%%%%%%%%%%%%%%%%%%%%%%%%%%%%%%%%%%%%%%%%%%%%%%%%%%%%%%%%%
\begin{center}
\begin{table}
\caption{Experimental  and theoretical structure parameters (z$_{Ti}$, z$_{O}$, z$_{Cl}$ are the c-axis coordinates
 in units of the c-axis length  and the distances Ti-O,Ti-Cl are given in  {\AA}).} 
\vspace{.3cm}
\begin{tabular}{c c c c c c c }
\hline
\hline
      \hspace{1cm}   & z$_{Ti}$   & z$_{O}$  & z$_{Cl}$  & Ti-O  & Ti-O-Ti  & Ti-Cl    \\
\hline
 Exp. \hspace{1cm}  &  0.1194    & 0.0551   & 0.3318    &1.96          &150$^{\circ}$       &  2.40   \\
 LDA  \hspace{1cm}  &  0.0846    & 0.0724   & 0.2999    &1.89          &174$^{\circ}$      &  2.41   \\ 
 LDA+U\hspace{1cm}  &  0.1133    & 0.0523   & 0.3223    &1.96          &151$^{\circ}$       &  2.38     \\ 
\hline 
\hline
\end{tabular}
\label{equipos}
\end{table}
\end{center}
%%%%%%%%%%%%%%%%%%%%%%%%%%%%%%%%%%%%%%%%%%%%%%%%%%%%%%%%%%

\vspace*{-0.99cm}

In Table \ref{equipos} we  show the equilibrium fractional c-axis coordinates, main distances and 
bonding angle of the three atoms obtained for the optimized lattice parameters. 
The most important difference with respect to the experimental structure is
 the shape of the Ti-O chain along the a-axis which in LDA
has become almost a linear chain 
(as shown also by the bonding angle Ti-O-Ti and the contraction
of the Ti-O distance).

\begin{figure}
\includegraphics[width=6.5cm]{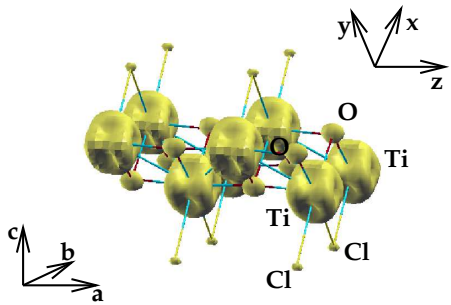}
\caption{Electron density surface for an isovalue of  $0.1 e{\AA}^3$  for the LDA equilibrium structure
(also shown the global and the local systems of reference).}
\label{3DLDA}
\end{figure}

Recalling that the local Titanium coordinate frame has the $z$ axis
parallel to $a$ and the $x$ and $y$ axes rotated by $45^{\circ}$  respect to $b$ and
$c$ (see Fig. \ref{3DLDA}),  
LDA provides  a t$_{2g}$ ground state where the   $d_{xy}$ and 
$d_{xz},d_{yz}$ contributions are  comparable in weight (0.2 and 0.27 electrons/eV respectively).  On the
other hand, the {\it ab initio} calculations\cite{saha_dasgupta_04} using the 
experimental structure indicate that the $d_{xy}$ weight is dominant .
The approximate degeneracy of the three t$_{2g}$ orbitals is  clearly shown in Fig.\ \ref{3DLDA} where 
the electron density surface in the energy range -1 eV to E$_F$ for an isovalue 
of $0.1 e/{\AA}^3$ is drawn and reflects the typical t$_{2g}$ spatial symmetry
with the minima of the isodensity along the octahedron axis.
The shortening of the Ti-O distance causes
 the local octahedron to be more regular
and the shape
of the surface around the titanium atom to be  an equal admixture of  
$d_{xy},d_{xz}$ and $d_{yz}$ orbitals. 
%%%%%%%%%%%%%%%%%%%%%%%%%%%%%%%%%%%%%%%%%%%%%%%
   
%%%%%%%%%%%%%%%%%%%%%%%%%%%%%%%%%%%%%%%%%%%%%%%%%%%%%%%%%%
Turning back to Table \ref{equipos}  we observe that while the Ti-Cl distance is in  good agreement
with  the experimental one,   the Ti-O distance  is   badly described
by the LDA.
We ascribe this  failure of LDA 
  to the improper description of electron correlations in titanium $d$-states 
  and we  show that  inclusion of $d$-orbital correlations on a mean-field level (LDA+$U$) causes the LDA
equilibrium structure to be unstable with respect to  the experimental one.
%%%%%%%%%%%%%%%%%%%%%%%%%%%%%%%%%%%%%%%%%%%%%%%%%%%%%%%%%

\begin{figure}
\vspace*{0.cm}
\includegraphics[width=7cm,keepaspectratio]{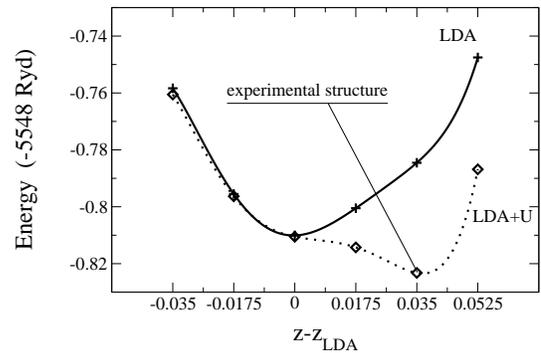}
\caption{LDA (solid line) and LDA+$U$ (dashed line) energy curves along the line (z,-z/2,z) in the space
 (z$_{Ti}$,z$_{O}$,z$_{Cl}$) which connects the LDA equilibrium structure
(taken as reference) to the experimental one. }
\label{enestruc}
\end{figure}

 In Fig.\ \ref{enestruc} we present the energy hypersurface 
 in the fractional coordinate space (z$_{Ti}$,z$_{O}$,z$_{Cl}$)  along the line connecting the LDA equilibrium 
 position and the experimental one which is represented approximately 
 by the vector (z,-z/2,z).
 
 It is clear that the inclusion of electron correlations distorts the LDA 
 energy line in a sort of  "double-well" curve whose minimum  is now very  close to
  the experimental one (see Fig.\ \ref{enestruc}, dotted line).
  Negative corrections to LDA total energy
  are expected for partially  orbitally polarized LDA groundstates, the amount
  of which depends
  on the correlation strength and  on the spatial extension of the $d$ orbitals 
  (namely, the U value and the titanium muffin radius). In fact, as shown in Fig\
   \ref{enestruc}, the two energy curves match at the LDA equilibrium value of $z$ where
    an orbitally non-polarized  ground  state  is found.
%%%%%%%%%%%%%%%%%%%%%%%%%%%%%%%%%%%%%%%%%%%%%%%%%%%%%%%%

As a final step in the structure determination of TiOCl
we want to find out the equilibrium structure according to  the  LDA+$U$ functional.
To determine the global minimum of the energy hypersurface   
we adopt a trial and error approach by 
performing  total energy calculations of all possible distorted 
structures  
around the experimental one and obtaining the minimum by a fitting procedure.

\begin{figure}
\includegraphics[width=5cm,keepaspectratio]{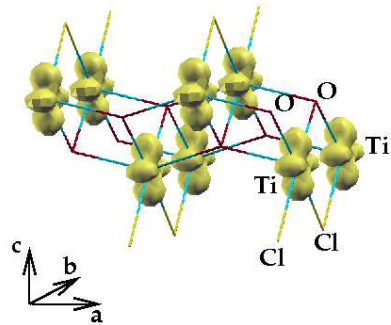}
\caption{Electron density surface for an isovalue of  $0.1 e/{\AA}^3$  for the LDA+$U$ equilibrium structure.}
\label{3DLDAU}
\end{figure}

As shown in Table \ref{equipos}, the experimental distances and bonding angles are well reproduced 
by the LDA+$U$ method; the disagreement in this case concerns the absolute fractional coordinates 
and it results, within every bilayer, in  a small reduction of the distance between the two Ti-O layers
and accordingly between the outer Cl ones. In Fig.\ \ref{3DLDAU} we show 
electron density plot of the LDA+$U$ equilibrium structure at the isosurface $0.1 e/{\AA}^3$.

 Preliminary calculations with the Generalized Gradient Approximation\cite{Perdew_96} (GGA) method
 show that the optimal equilibrium structure is improved with respect to the LDA results
 but it is still far from the experimental one.

%%%%%%%%%%%%%%%%%%%%%%%%%%%%
%%%%%%%%%%%%%%%%%%%%%%%%%%%%% 

{\it Phonons within LDA and LDA+$U$}.-
The phonon frequencies  within the LDA approximation were calculated by considering the 
forces acting on the three atoms when they were displaced with respect to the  equilibrium 
positions in all possible fashions (in-phase and out-of-phase  displacements of one, two and three atoms).
The  force  values were fitted to  linear polynomials in the  c-axis coordinates
and were determined within the uncertainty of  1 mRy/a.u..

\begin{center}
\begin{table}
\caption{LDA and LDA+$U$ phonon frequencies  compared to experimental data  
and mode assignment (relative atomic displacement, i.e. components of 
the eigenvectors divided by $\sqrt{M_i}$ and normalized
to the maximum amplitude).}

\begin{tabular}{c|c|c|c|c}
                                &   & $\beta$ & $\gamma$ & $\delta$  \\
\hline
\hline
    {\bf Expt.}  &  freq.(cm$^{-1})$   &     {\bf 203}      &   {\bf 365}     &    {\bf 430 } \\
\hline
\hline
   {\bf LDA  }   &   freq.(cm$^{-1})$   &    {\bf 154}      &  {\bf 351}     &   {\bf  424}      \\
 (mode assign.)  &    Ti                &     0.909    &  0.822    &    -0.025\\
                 &     O                &    -0.157    &  0.311    &     1.000   \\
                 &    Cl                &     1.000    &  -1.000   &     0.104 \\
\hline
\hline
{\bf LDA+$U$  }  &   freq.(cm$^{-1})$   &   {\bf  192}      & {\bf  336}     &   {\bf  407}      \\
 (mode assign.)  &    Ti                &     0.557    &   0.307    &   -0.775 \\
                 &     O                &     -0.048  &    1.000  &       1.000  \\
                 &    Cl                &      1.000  &     -0.213    &     0.614  \\
\hline
\hline
\label{eigenv}
\end{tabular}
\end{table}
\end{center}

\vspace*{-0.99cm}

Table  \ref{eigenv} displays the LDA frequencies and 
the corresponding relative atomic displacements in comparison with the Raman 
 frequencies\cite{lemmens_03_2}. 
 The lowest frequency mode is a titanium-chlorine 
 in-phase and oxygen out-of-phase mode and it shows a considerable 
 disagreement with respect to the
 experimental  value.
The second mode is a titanium-oxygen  in-phase and chlorine out-of-phase mode 
 and the third mode
 is mainly given by oxygen with small in-phase contribution of
 chlorine and almost negligible out-of-phase contribution from titanium.
To test the accuracy of the fitting polynomials we crosschecked a posteriori the frequency values by displacing  
the atoms along  the  eigenvector directions finding the larger deviation ($\sim 10$\%) for the lowest
mode.

Within the LDA+$U$ method we also calculated phonon frequencies and modes by means of a
quadratic fitting of total energy surface.
The frequencies and assignments are displayed in Table \ref{eigenv}.
The numerical error of the calculation is 10$^{-4}$ Ry  and  affects
mainly the
second frequency value. 
The error due to the fitting function proved to be negligible 
confirming the harmonic character of all the three vibrations. 
  In LDA+U the first frequency becomes considerably 
improved with respect to the LDA value and in good agreement
to the experimental mode. 
The second  mode upon inclusion of the numerical error is found in  good agreement with the experiment
and the third frequency disagrees by  about 4\%. 

Regarding the mode assignments in the two approaches, both LDA and
 LDA+$U$ calculations agree 
as far as the  in-phase and out-of-phase features are concerned.
From a quantitative point of view instead, LDA+$U$ associates  in contrast to LDA, a larger amplitude to chlorine
and smaller to oxygen  in the first mode;  
in the second mode the principal elongation
is  the oxygen one with smaller contributions from titanium and chlorine 
 while the latter play the major role in LDA and finally the third mode 
is characterized by roughly balanced amplitudes for all the three atoms 
 in LDA+$U$ while in LDA the mode is almost pure oxygen oscillation.
 Calculation of the  Raman cross sections for the three modes
 would bring to evidence 
  the main differences between the  LDA and LDA+$U$  atomic displacements
 which can be tested against the experimental Raman intensities.
 
\vspace{0.cm}
{\it Conclusions}.-
Summarizing, we have presented  DFT calculations for the
 $A_g$ Raman phonon modes in TiOCl within LDA and LDA+$U$. Within LDA we obtain
 that the optimized equilibrium structure deviates considerably respect
to the experimental one  while  consideration of electron correlation within the 
LDA+$U$ approach brings the DFT minimum very near to the experimental
 data (importance of correlation effects has been already pointed out by P. Labeguerie {\it et  al.}\cite{Labeguerie_05} and by M. Merawa {\it et al.}\cite{Merawa_03}).
 In both approaches, by calculation of the dynamical matrix, we
 obtain the three $A_g$ modes in the $Pmmn$ symmetry.
  While both LDA and LDA+$U$ show an overall qualitative agreement with the results from Raman scattering experiments,
 quantitatively  LDA+U produces for the lower mode
 the best agreement with the experiment while LDA performs better for the second and third mode.

{\it Acknowledgments}.-
 One of us (L.P.)   thanks the WIEN2k-users-web and P. Blaha for useful
 comments regarding the WIEN2k code and T. Kokalj for providing the graphic code
 XCRYSDEN. We also thank the German Science Foundation for financial support.

%%%%%%%%%%%%%%%%%%%%%%%%%%%%%%%%%%%%%%%%%%%%%%%%%%%%%%%%%%%%
%%%%%%%%%%%%%%%%%%%%%%%%%%%%%%%%%%%%%%%%%%%%%%%%%%%%%%%%%%%%

%%%%%%%%%%%%%%%%%%%%%%%%%%%%%%%%%%%%%%%%%%%%%%%%%%%%%%%%%%%%%%%%%%%%%%%%%
%%%%%%%%%%%%%%%%%%%%%%%%%%%%%%%%%%%%%%%%%
%%%%%%%%%%%%%%%%%%%%%%%%%%%%%%%%%%%%%%%%%%%%%%%%%%%%%%%%%%%%%%

\end{document}